\documentclass{article}

\usepackage{float}
\usepackage{amsmath}
\usepackage{graphicx}
\usepackage{makecell}
\usepackage[hidelinks]{hyperref}
\usepackage{algorithm}
\usepackage{algpseudocode}
\usepackage{amssymb}
\usepackage{authblk}
\usepackage{booktabs}
\usepackage[small]{titlesec}
\usepackage[OT1]{fontenc}
\usepackage[style=numeric,natbib=true,backend=bibtex,url=false,mincitenames=1,
            maxcitenames=2]{biblatex}
\algrenewcommand\alglinenumber[1]{\footnotesize #1:}
\algrenewcommand\algorithmicrequire{\textbf{Input:}}
\algrenewcommand\algorithmicensure{\textbf{Output:}}
\makeatletter
\renewcommand{\ALG@beginalgorithmic}{\small}
\makeatother

\AtEveryBibitem{%
  \clearfield{issn}%
}

\title{
    \Large\textbf{Single-Asset Adaptive Leveraged Volatility Control}
}
\author[1]{Nikhil Devanathan\thanks{email: nikhil.devanathan@blackrock.com}}
\author[1]{Dylan Rueter}
\author[1]{Stephen Boyd}
\author[1]{\authorcr Emmanuel Cand\`es}
\author[1]{Trevor Hastie}
\author[1]{Mykel J. Kochenderfer}
\author[2]{\authorcr Arpit Apoorv}
\author[2]{David Soronow}
\author[2]{Igor Zamkovsky}

\affil[1]{BlackRock AI Lab}
\affil[2]{BlackRock Index Services}
\date{\today}

\begin{document}
    \maketitle

\begin{abstract}
This paper introduces a methodology for constructing a market index composed of a liquid risky asset and a liquid risk-free asset that achieves a fixed target volatility. Existing volatility-targeting strategies typically scale portfolio exposure inversely with a variance forecast, but such open-loop approaches suffer from high turnover, leverage spikes, and sensitivity to estimation error---issues that limit practical adoption in index construction. We propose a proportional-control approach for setting the index weights that explicitly corrects tracking error through feedback. The method requires only a few interpretable parameters, making it transparent and practical for index construction. We demonstrate in simulation that this approach is more effective at consistently achieving the target volatility than the open-loop alternative.
\end{abstract}

\section{Introduction}

We address the problem of constructing a financial index---a rule-based portfolio---from two assets: a liquid risky asset (e.g., an equity fund) and a liquid risk-free asset (e.g., cash or short-term government bonds). The objective is for the index to realize a pre-specified target volatility, where volatility refers to the standard deviation of returns. We operate under two constraints: the risky asset cannot be shorted, and there is an upper bound on leverage, i.e., the degree to which capital is borrowed to amplify exposure to the risky asset.

From a control-theoretic perspective, this is a setpoint tracking problem with input saturation: the ``plant'' is the stochastic return process of the risky asset, the control input is the portfolio weight, and the regulated output is the realized volatility of the index. The saturation arises from the leverage limit, which imposes a hard upper bound on the control authority.

Volatility-controlled indices are widely used in institutional investment, structured products, and risk-managed strategies. By maintaining a stable risk profile, such indices allow investors to separate asset selection decisions from risk-level decisions. From an engineering standpoint, the problem is appealing because it requires regulating a noisy, non-stationary process using simple, interpretable rules---a setting where feedback control has clear advantages over open-loop strategies.

The core difficulty is that asset volatility is latent and time-varying. Unlike a physical system where sensor noise corrupts a measurable state, here the ``true'' volatility is never directly observed---only noisy return realizations are available. Volatility estimates derived from historical returns are inherently lagged and uncertain. Moreover, financial volatility exhibits well-documented stylized facts: it clusters (periods of high volatility tend to persist), it responds asymmetrically to positive and negative returns, and it occasionally undergoes abrupt regime shifts during market crises \cite{hu2018, xunyu2002}. These properties cause open-loop strategies that rely solely on volatility forecasts to systematically lag behind the true volatility, resulting in persistent tracking errors.

The standard approach to volatility targeting, formalized by Moreira and Muir \cite{moreira2016}, scales portfolio exposure inversely with a forecast of variance. Large empirical studies confirm that such open-loop rules can improve Sharpe ratios and reduce tail risk across many asset classes \cite{harvey2018}. However, these same studies highlight practical issues: high portfolio turnover, leverage spikes during volatility transitions, and sensitivity to the choice of volatility estimator.

Several refinements have been proposed. Conditional schemes restrict scaling to extreme volatility regimes \cite{bongaerts2020}, while smoothing techniques based on exponentially weighted moving averages (EWMAs) or Bayesian updating dampen weight oscillations \cite{bernardi2022}. Despite these improvements, empirical work has shown that out-of-sample benefits can vanish once implementation frictions such as transaction costs and rebalancing delays are considered \cite{cederburg2020, duncan2008}. Crucially, all of these approaches remain fundamentally open-loop: they set portfolio weights based on a volatility forecast without incorporating feedback from the realized volatility of the portfolio itself.

The control-systems literature has applied feedback principles to related financial problems---for example, prior work on stock trading has used proportional and PI controllers to dynamically adjust position sizes \cite{malekpour2013, malekpour2016}, and earlier work has addressed volatility tracking in stock markets \cite{cvitanic2000}---but a gap remains: there is no simple, interpretable closed-loop controller specifically designed for volatility targeting under leverage constraints. Existing financial methods are open-loop, and existing control-theoretic work addresses different objectives (e.g., maximizing growth, minimizing drawdown) rather than tracking a volatility setpoint.

We cast volatility targeting as a proportional feedback-control problem. The controller observes the realized volatility of the index (estimated through a EWMA of squared returns), computes the tracking error relative to the target, and adjusts the portfolio weight accordingly. The EWMA serves as a simple state observer, analogous to a low-pass filter that extracts the volatility signal from noisy return data. We use clipping to enforce the leverage constraint, a technique with well-established properties for constrained control \cite{kvasnica2012, gilbert1991}, and define the tracking error in logarithmic space for scale invariance. The controller includes a smoothing term that acts as a lag filter on the control signal, preventing excessive reactivity to transient fluctuations.

A key feature of our approach is its deliberate simplicity. The volatility controller requires only four interpretable parameters: a proportional gain, upper and lower bounds on the control signal, and a smoothing factor. This parsimony is a design choice: in index construction, transparency and auditability are paramount, and complex black-box methods face significant barriers to adoption. We validate the method through parameter sensitivity analysis showing robustness across a wide range of gain and smoothing values, and through temporally out-of-sample testing on a diverse cohort of ETFs demonstrating generalization beyond the tuning asset.

\section{Methodology}
Our goal is to construct an index from a single risky asset and a single risk-free asset (for illustration, we use cash) which achieves a specified level of volatility $\sigma^{\text{tar}} > 0$ without exceeding a limit $L>0$ on the weight allocated to the risky asset. We call $\sigma^{\text{tar}}$ the target volatility and $L$ the leverage limit. We denote by $t_1,t_2, \ldots$ the time sequence at which we
observe the returns of the risky asset, realize interest on cash balances, and have the opportunity to rebalance
the index. We define the trailing returns of the risky asset and risk-free asset at time $t_k$ as $r_k, r^\text{rf}_k\in \mathbb{R}$ respectively.

\paragraph{Index weights and returns} An index is a pair of weights $(w_k,c_k)$ at time $t_k$ such that $w_k\in \mathbb{R}$ is the weight of the risky asset and $c_k\in \mathbb{R}$ that of cash. The weights must satisfy the conditions
\[
w_k+c_k=1, \qquad 0 \leq w_k\leq L.
\]
The weights represent the fraction of the index invested in the risky asset and cash, respectively. A negative cash weight corresponds to leverage, or borrowing cash to invest more in the risky asset, which we can take in a limited capacity. When taking leverage, we pay interest on the borrowed cash rather than accruing positive interest on the cash. Between two consecutive times $t_{k-1}$ and $t_k$, the index return $r^\text{ind}_k$ is given by
\[r^\text{ind}_k=r_kw_{k-1} + r^\text{rf}_kc_{k-1}.\]
\paragraph{Realized volatility} The realized volatility $\hat \sigma_k$ of the risky asset at time $t_k$ is defined as the square root of an EWMA of the squared risky asset returns $r_1,\ldots,r_k$,
\begin{equation}\label{e-ewma}
\left(\hat \sigma_k \right)^2= \frac{1-\beta}{1-\beta^k}\sum_{j=1}^k\beta^{k-j}\left(r_j\right)^2,
\end{equation}
with a halflife $h>0$ and decay factor $\beta=\exp(-\log(2)/h)$ \cite{johansson2023}. (This is an EWMA estimate of the second moment, but we use the fact that daily returns have means that are much smaller in magnitude than the returns.) Note that $\hat \sigma_k$ is known at the time that $w_k$ is decided.

\paragraph{Realized index volatility} Analogously, we define the realized index volatility $\hat \sigma^\text{ind}_k$ at time $t_k$ as the square root of the EWMA of the squared index returns $r_1^\text{ind},\ldots,r_k^\text{ind}$,
\begin{equation}\label{e-ind-ewma}
\left(\hat \sigma^\text{ind}_k \right)^2= \frac{1-\beta}{1-\beta^k}\sum_{j=1}^k\beta^{k-j}\left(r_j^\text{ind}\right)^2.
\end{equation}
Note that $\hat \sigma^\text{ind}_k$ is known at the time that $w_k$ is decided.

\paragraph{Tracking error} We define our relative instantaneous volatility tracking error $e_k$ at time $t_k$ as
\begin{equation}\label{e-vol-error}
    e_k = \log(\hat{\sigma}^\text{ind}_k/\sigma^{\text{tar}}).
\end{equation}
The instantaneous tracking error $e_k$ is approximately the percentage difference between the realized index volatility $\hat{\sigma}^\text{ind}_k$ and the target volatility $\sigma^{\text{tar}}$. Over a period $t_j,t_{j+1},\ldots,t_k$, the average absolute volatility tracking error is
\[
    \frac{1}{k-j+1}\sum_{i=j}^k |e_i|.
\]

\paragraph{The volatility control problem} The problem is to determine a policy that chooses $w_k$ and $c_k$ at time $t_k$ based on information known at time $t_k$ (such as $\hat \sigma_k, \hat\sigma_k^\text{ind}, e_k$) in such a way that the average absolute volatility tracking error is minimized.

\subsection{Open-loop approach}\label{s-open-loop}

In the ideal case, the returns of the risk-free asset are known constants and the returns of the risky asset are independently and identically distributed according to a normal distribution with mean $\mu$ and standard deviation $\sigma>0$ at each time $t_j$. In this case, the index returns at time $t_{k+1}$ are distributed according to a normal distribution with mean $w_k\mu + c_k r^\text{rf}_{k+1}$ and standard deviation $w_k\sigma$. If we assume that $|\mu|\ll \sigma$ and $r^\text{rf}_k\ll \sigma$, then the expected value of $(r_{k+1})^2$ is approximately $\sigma^2$ and the expected value of $(r_{k+1}^\text{ind})^2$ is approximately $w_k^2\sigma^2$. It follows by the linearity of expectation that the estimated variance $(\hat \sigma_k)^2$ of the risky asset should be approximately $(\sigma)^2$.

In this case, a natural approach to choosing $w_k$ and $c_k$ such that the expected value of $(r_{k+1}^\text{ind})^2$ is equal to $\sigma^{\text{tar}}$ is to set
\begin{equation}\label{e-naive-sol}
w_k=\min\{\sigma^{\text{tar}}/\hat\sigma_k,L\}, \qquad c_k = 1 - w_k.
\end{equation}
When $\sigma^{\text{tar}}/\hat\sigma_k\leq L$, we set the risky asset weight $w_k$ so that the expected value of the index variance is equal to the target variance. When $\sigma^{\text{tar}}/\hat\sigma_k > L$, we set $w_k=L$, accepting that this is the best we can do given the leverage limit. We call this approach the ``open-loop'' approach because the weights $w_k$ are set without any regard to the previous index weights $w_1,\ldots,w_{k-1}$.

This methodology is summarized in Algorithm \ref{a-naive}.
\begin{algorithm}
    \caption{Open-loop solution to index construction}
    \label{a-naive}
    \begin{algorithmic}[1]
        \State \textbf{Input:} $\sigma^{\text{tar}}, L$
        \For {$k = 1,2,\ldots$}
            \State Observe risky asset volatility $\hat{\sigma}_k$
            \State Set risky asset weight $w_k=\text{min}\{\sigma^{\text{tar}}/\hat{\sigma}_k,L\}$
            \State Set cash weight $c_k=1 - w_k$
        \EndFor
    \end{algorithmic}
\end{algorithm}

In practice, although this method may sometimes perform reasonably, it fares poorly when there are
changes in the prevailing volatility of the risky asset. This is illustrated in \S\ref{ap-result}.

\subsection{Volatility control}
We now modify the methodology from \S\ref{s-open-loop} to better react to shifts in the volatility
of the risky asset. To achieve this, we add a \textit{control parameter} $\kappa_k\in \mathbb{R}$ to our policy \eqref{e-naive-sol} and set $w_k$ and $c_k$ such that
\begin{equation}\label{e-vol-sol}
\begin{split}
    w_k &= \min\{\exp(\kappa_k)\sigma^{\text{tar}}/\hat \sigma_k,L\} \\
    c_k &= 1 - w_k.
\end{split}
\end{equation}

The $\exp(\kappa_k)$ scales the risk target and thus allows
us to adjust for persistent overestimation or underestimation of the volatility of the risky asset. Operating in logarithmic space is motivated by the empirical observation that volatility varies multiplicatively: for an asset with a mean volatility of $X\%$, periods of $2X\%$ and $\frac{1}{2}X\%$ volatility are approximately equally likely. This choice also aligns with prior work on log-optimal portfolios, which has established the benefits of logarithmic formulations for scale-invariant control \cite{proskurnikov2023, hsieh2020}.

It remains to discuss how to dynamically set the control parameter $\kappa_k$, and we describe a simple methodology, which depends on only four parameters. The four parameters are a proportional gain $g>0$, a control parameter lower bound $\kappa_\text{min}<0$, an upper bound $\kappa_\text{max}>0$, and a smoothing factor $\theta\in (0,1)$. After setting these parameters, we compute $\kappa_k$ via
\begin{equation}\label{e-kappa}
    \kappa_k = (1-\theta)\text{clip}(-g e_k; I) + \theta \kappa_{k-1}
\end{equation}
in which
\begin{equation}
\label{e-terms}
I = [\kappa_\text{min}, \kappa_\text{max}], \qquad \text{clip}(t; [a,b]) = \begin{cases} a, & t < a,\\
    t, & a\le t \le b,\\
    b & t> b.
\end{cases}
\end{equation}

This controller is designed to adjust $\kappa_k$ based on the discrepancy between the realized volatility of the index $\hat{\sigma}^\text{ind}_k$ and the target volatility $\sigma^{\text{tar}}$. To simplify, assume $\theta = 0$ (no smoothing) and that the volatility of the index has recently been lower than the target. Then the error $e_k$ will be negative, and the term $-ge_k$ will be positive, meaning that we will target a higher volatility in the next iteration. Conversely, if the volatility of the index has recently been higher than the target, then $e_k$ will be positive, and $-ge_k$ will be negative, meaning that we will target a lower volatility in the next iteration.

The gain parameter $g$ influences how reactive $\kappa_k$ is to discrepancies between the realized and target volatilities. The clipping imposes reasonable limits on $\kappa_k$ to prevent it from becoming too large or too small. The smoothing factor $\theta$ determines how quickly $\kappa_k$ can change; a larger value of $\theta$ means that $\kappa_k$ will change more slowly, while a smaller value implies $\kappa_k$ will change more quickly.

Since the volatility control methodology assumes our index has some history (for volatility estimation),
we recommend either running or simulating the open-loop method until some time $t_\text{start}$ before switching to the volatility control
methodology.

The volatility control methodology is summarized in Algorithm \ref{a-vol} and its performance is demonstrated in \S\ref{ap-result}.
\begin{algorithm}
    \caption{Volatility control for index weight selection}
    \label{a-vol}
    \begin{algorithmic}[1]
        \State \textbf{Input:} $\sigma^{\text{tar}}, L, g, \kappa_\text{min}, \kappa_\text{max}, \theta$
        \State Set $\kappa_0 = 0$
        \For {$t_k = t_\text{start},\ldots$}
            \State Observe risky asset volatility $\hat{\sigma}_k$
            \State Produce an index volatility estimate $\hat{\sigma}^\text{ind}_k$ from historical index returns
            \State Compute volatility control error as in \eqref{e-vol-error}
            \State Compute volatility control parameter as in \eqref{e-kappa}
            \State Set risky asset weight $w_k=\text{min}\{\exp(\kappa_k)\sigma^{\text{tar}}/\hat{\sigma}_k,L\}$
            \State Set cash weight $c_k=1 -w_k$
        \EndFor
    \end{algorithmic}
\end{algorithm}

\section{Results}\label{ap-result}
In this section, we present results about the performance of Algorithms \ref{a-naive} and \ref{a-vol}. We first compare the methods on a single asset, then analyze parameter sensitivity, and finally evaluate generalization across a broad cohort of ETFs.

\subsection{Baseline Comparison}

\paragraph{Simulation details} Our simulation spans the trading days from June 8, 2000 to December 31, 2024, inclusive.
The risky asset is IVV, the iShares Core S\&P 500 exchange traded fund (ETF), which provides broad exposure to large-cap U.S. equities. We use close-to-close returns for IVV and assume that returns are observed, new weights are computed, and the index is rebalanced at market close. We model transaction costs as a 5 basis point bid-ask spread applied to each rebalancing trade. The return on cash is the daily Federal Funds Effective Rate. We compute the risky asset volatility estimate $\hat{\sigma}_k$ as the square root of an EWMA of the squared risky asset returns $r_1,\ldots,r_k$ with a halflife of 126 trading days.
We set the leverage limit to $L=1.5$ and the target volatility to $\sigma^{\text{tar}}=0.15/\sqrt{252}$, corresponding to an annualized volatility of $15\%$.

In Algorithm \ref{a-vol}, we set $g=55$, $\kappa_\text{min}=-1$, $\kappa_\text{max}=1$, and $\theta=0.6$; the selection of these parameters is discussed in \S\ref{s-sensitivity}.
We run the method for the first 10 trading days before engaging control.
We compute the index volatility estimate $\hat{\sigma}^\text{ind}_k$ as the square root of an EWMA of the
squared index returns $r_1^{\text{ind}},\ldots,r_k^{\text{ind}}$ with a halflife of 126 trading days.

\begin{figure}[H]
\centering
\includegraphics[width=\linewidth]{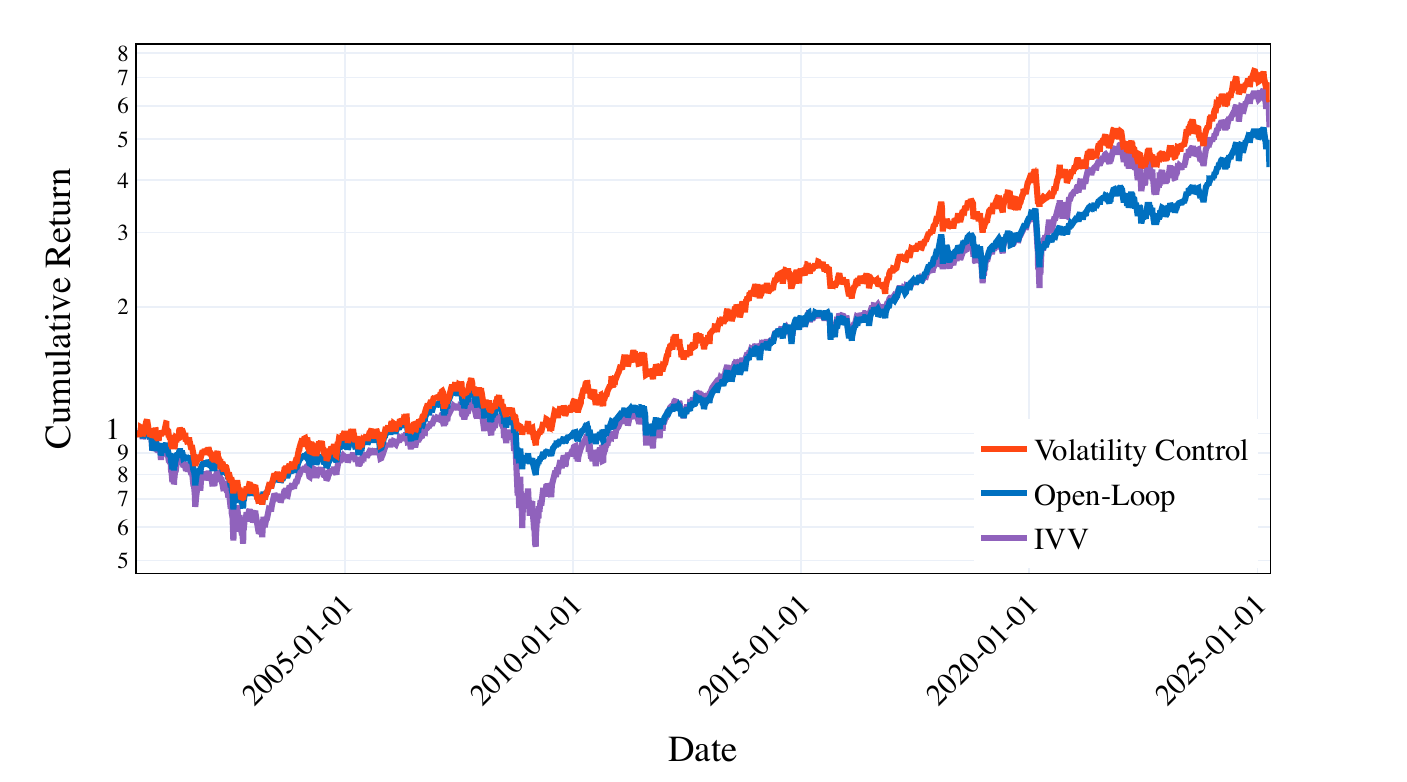}
\caption{Cumulative returns of three indices.}
\label{f-return}
\end{figure}

\begin{figure}[H]
\centering
\includegraphics[width=\linewidth]{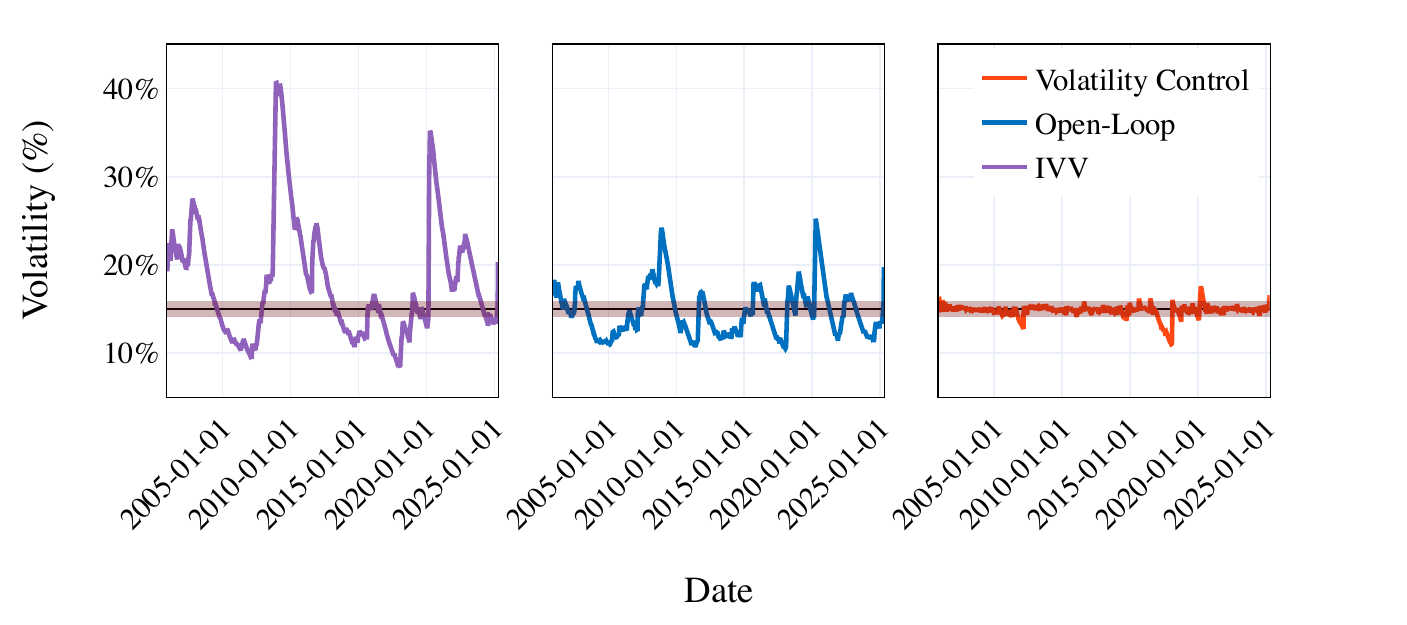}
\caption{Running annualized estimated volatility of three indices. 15\% target volatility is shown as a solid black line. A 90\% confidence interval for the running volatility estimate of an asset with a true volatility of $15\%$ is shown as a shaded region.}
\label{f-vol}
\end{figure}

\paragraph{Discussion} Figure \ref{f-return} shows the cumulative returns of IVV, the open-loop method, and the volatility control method. Figure \ref{f-vol} shows the running estimated volatility of IVV and that of the two
methods over the same period. The Monte Carlo methodology described in Appendix \ref{ap-eval} was used to compute 10th and 90th percentiles for
the running volatility estimate of an asset with a true volatility of $15\%$.

Table \ref{t-results} summarizes the performance of the three methods over the simulation period. The volatility tracking error is computed as the
mean absolute error between the running volatility estimate of the method and the target volatility $\sigma^{\text{tar}}$, scaled by $\sqrt{252}$ to
give an annualized value. The Sharpe ratio is the annualized return in excess of the Fed Funds Effective Rate divided by annualized volatility, the Kalmar ratio is the annualized return divided by maximum drawdown, and turnover is the sum of absolute weight changes over the period. Although IVV does not track a volatility target, we still provide a volatility tracking error for IVV for reference.

As shown in the figures and table, the volatility control method outperforms the open-loop method in terms of return, volatility tracking error, and maximum drawdown.

\begin{table}[H]
\centering
\caption{Summary of performance metrics for three indices.}
\label{t-results}
\begin{tabular}{@{}lrrr@{}}
\toprule
                           & IVV    & Open-Loop & Volatility Control \\
\midrule
Volatility Tracking Error  & $5.2\%$   & $2.3\%$              & $\mathbf{0.4\%}$ \\
Annualized Return          & $7.8\%$   & $6.8\%$              & $\mathbf{8.2\%}$ \\
Annualized Volatility      & $19.1\%$  & $\mathbf{14.9\%}$             & $\mathbf{14.9\%}$ \\
Sharpe Ratio               & $0.31$    & $0.33$               & $\mathbf{0.42}$ \\
Kalmar Ratio               & $0.14$    & $0.18$               & $\mathbf{0.22}$ \\
Maximum Drawdown           & $55.3\%$  & $38.6\%$             & $\mathbf{37.1\%}$ \\
Turnover                   & $\mathbf{0\%}$  & $93\%$             & $1105\%$ \\

\bottomrule
\end{tabular}
\end{table}

\subsection{Parameter Sensitivity}\label{s-sensitivity}

The volatility controller has two key tunable parameters: the proportional gain $g$ and the smoothing factor $\theta$. We fix $\kappa_\text{min}=-1$ and $\kappa_\text{max}=1$ throughout, as these bounds primarily serve as safety limits. To understand sensitivity to $g$ and $\theta$, we conduct a grid search over $g\in\{0, e^0, e^{0.5}, e^1, \ldots, e^5\}$ (12 values) and $\theta\in\{0, 0.1, \ldots, 0.9\}$ (10 values). For each parameter combination, we simulate the volatility controller on IVV over the period June 8, 2000 to December 31, 2009 and compute three metrics: volatility tracking error, change in Kalmar ratio relative to IVV, and portfolio turnover.

\begin{figure}[H]
\centering
\includegraphics[width=\linewidth, interpolate=false]{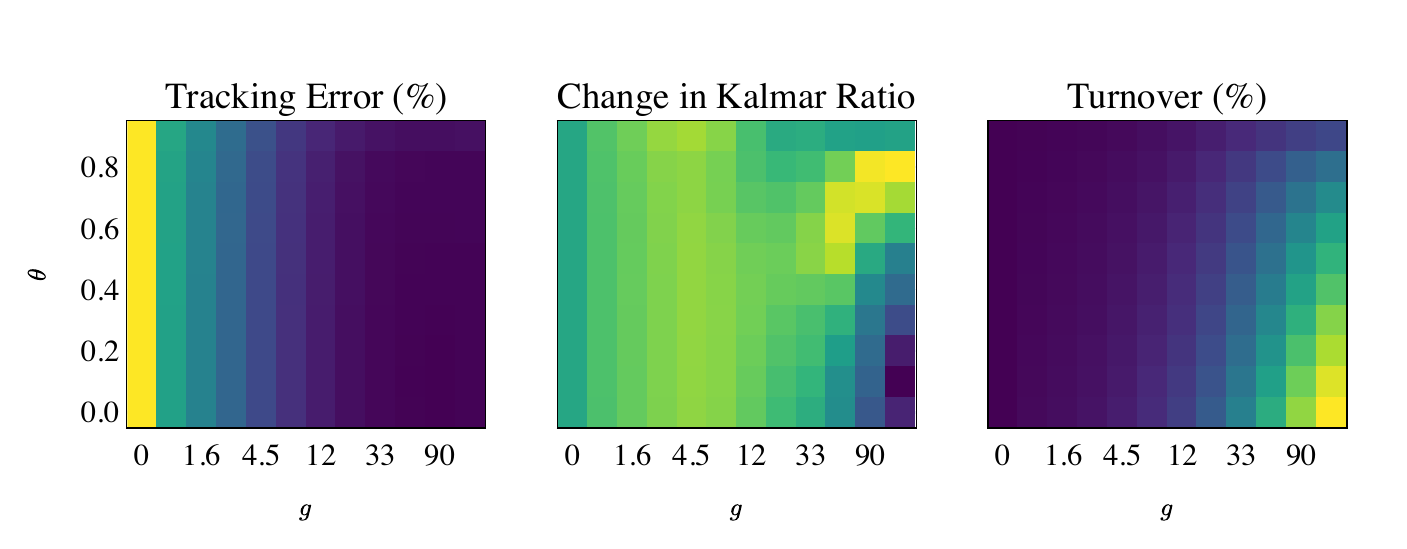}
\caption{Parameter sensitivity of the volatility controller on IVV (June 2000--December 2009). Left: volatility tracking error (0.29\%--2.52\%). Center: change in Kalmar ratio relative to the underlying asset ($-0.056$ to $+0.066$). Right: portfolio turnover (108\%--3558\%). Lighter colors indicate higher values. All three metrics vary smoothly over the parameter space, indicating robustness to parameter choice.}
\label{f-heatmap}
\end{figure}

Figure \ref{f-heatmap} shows heatmaps of these metrics over the $(g, \theta)$ grid. Tracking error decreases monotonically in $g$ and is largely insensitive to $\theta$. Change in Kalmar ratio is positive for nearly all parameter combinations, with a ridge of particularly favorable values between $(g, \theta) = (148, 0.8)$ and $(55, 0.6)$. Turnover increases monotonically in $g$ and decreases monotonically in $\theta$.

These surfaces suggest that many parameter combinations achieve good performance. Higher gain reduces tracking error but increases turnover; the smoothing factor $\theta$ effectively mitigates turnover while preserving the low tracking error achieved by high gain. Based on this analysis, reasonable choices include $(g, \theta) \in \{(55, 0.6), (90, 0.8)\}$. For both the baseline comparison in \S3.1 and out-of-sample testing, we select $(g, \theta) = (55, 0.6)$ from within this robust region.

\subsection{Multi-Asset Evaluation}

The preceding analysis establishes that the volatility controller is robust to parameter choice on a single asset. We now assess whether the method generalizes to assets not used for tuning. We evaluate both methods on a broad cohort of ETFs using a temporally out-of-sample methodology: parameters are selected using IVV data from June 2000 to December 2009 (the period analyzed in \S3.2), and we then test on a diverse set of ETFs over the disjoint period January 1, 2010 to December 31, 2024. We use the parameters $(g, \theta) = (55, 0.6)$ identified above, applied uniformly across all assets without per-asset tuning. As in the baseline comparison, we include a 5 basis point bid-ask spread.

The ETF cohort consists of all ETFs in our dataset with positive price data throughout the test period and annualized returns exceeding 5\%. This filter yields a diverse set of 44 ETFs spanning U.S. and international equities, commodities, and sector funds: ACWI, EFA, EWD, EWJ, EWK, EWL, EWN, EWT, IAI, IAK, IAT, IAU, IBB, IDU, IEO, IGM, IGV, IHF, IHI, IJH, IJR, IOO, ITA, ITB, IVV, IWB, IWD, IWF, IWM, IWN, IWO, IWV, IXN, IYC, IYF, IYH, IYJ, IYK, IYM, IYR, IYW, OEF, SOXX, and THD.

\begin{figure}[H]
\centering
\includegraphics[width=\linewidth]{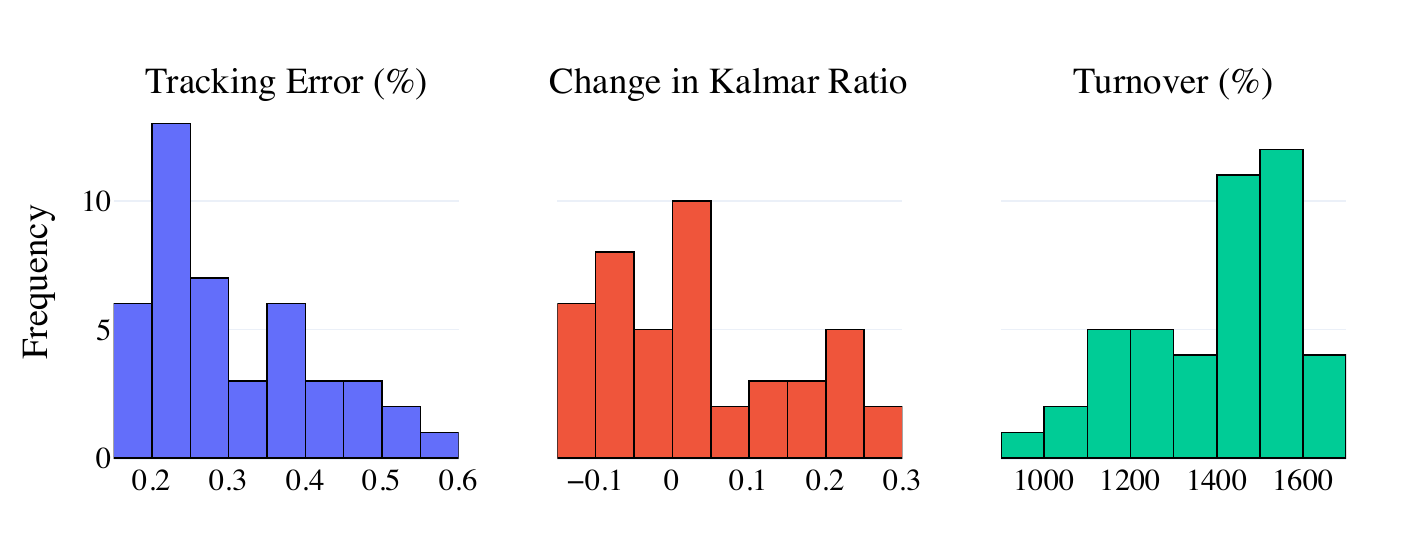}
\caption{Distribution of performance metrics for the volatility controller across an ETF cohort (January 2010--December 2024). Left: volatility tracking error. Center: change in Kalmar ratio relative to the underlying asset. Right: portfolio turnover.}
\label{f-histogram-vc}
\end{figure}

\begin{figure}[H]
\centering
\includegraphics[width=\linewidth]{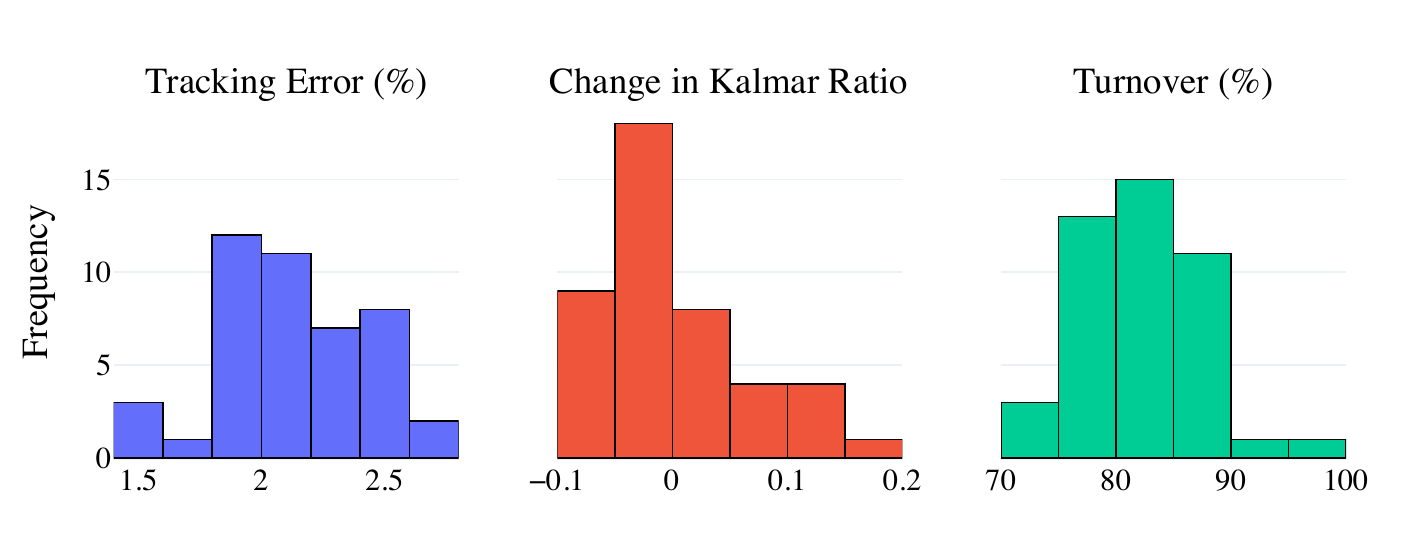}
\caption{Distribution of performance metrics for the open-loop method across the same ETF cohort. Left: volatility tracking error. Center: change in Kalmar ratio relative to the underlying asset. Right: portfolio turnover.}
\label{f-histogram-ol}
\end{figure}

Figures \ref{f-histogram-vc} and \ref{f-histogram-ol} show histograms of three metrics across the ETF cohort for the volatility controller and open-loop method, respectively. For the volatility controller, tracking error is predominantly below 0.4\% and entirely below 0.6\%, demonstrating that the controller achieves its core objective across diverse assets. The change in Kalmar ratio is positive for the majority of ETFs, with values as high as 0.25, though a minority exhibit negative values---none below $-0.15$. Turnover ranges from 800\% to 1800\%.

For the open-loop method, tracking error ranges from 1\% to 3\%, consistent with the IVV baseline. The change in Kalmar ratio is similar in shape but smaller in magnitude ($-0.1$ to $0.2$), with most of the mass slightly below zero. Turnover is substantially lower (70\%--100\%), reflecting the method's less reactive nature.

Both methods generalize consistently from IVV to the broader ETF cohort: the volatility controller achieves significantly lower tracking error at the cost of higher turnover, while the open-loop method offers lower turnover but substantially worse volatility tracking. The high turnover of the volatility controller is appropriate for highly liquid assets where rebalancing costs are minimal; for less liquid assets, the open-loop method may be preferable despite its weaker tracking performance.

\section*{Acknowledgments}
We would like to acknowledge Gabriel Maher, Rob Tibshirani, and Ludovic Breger for their work on an initial methodology we built upon. We would additionally like to thank Veronica Mai, Logan Bell, Alex Tzikas, Raphael Chinchilla, and Ron Kahn for helpful discussions and feedback.

\printbibliography

\appendix
\section{Uncertainty Quantification for Volatility Estimates}\label{ap-eval}
Suppose we have constructed an index with returns $r_1^\text{ind}, r_2^\text{ind}, \ldots$ and we estimate its volatility using some method (e.g., the EWMA defined in \eqref{e-ind-ewma}) to obtain estimates $\sigma^\text{eval}_1, \sigma^\text{eval}_2, \ldots$. Even if the index had a volatility exactly equal to the target $\sigma^{\text{tar}}$, any finite-sample estimate $\sigma^\text{eval}_k$ will almost surely differ from $\sigma^{\text{tar}}$ due to sampling variability. To evaluate whether our index is successfully tracking the target, we must understand how much deviation to expect from a perfectly tracking index.

We propose the following Monte Carlo methodology to quantify this uncertainty:
\begin{enumerate}
    \item Sample $N$ synthetic return values $r_1^\text{synth},\ldots,r_N^\text{synth}$ independently from a normal distribution with mean $0$ and standard deviation $\sigma^{\text{tar}}$.
    \item Compute volatility estimates $\sigma^\text{synth}_1, \ldots, \sigma^\text{synth}_N$ from the synthetic returns using the same estimation method used to compute $\sigma^\text{eval}_k$ from the index returns.
    \item Discard the first $n$ values (a burn-in period) and compute statistics of interest from $\sigma^\text{synth}_{n+1},\ldots,\sigma^\text{synth}_N$.
\end{enumerate}
Because the synthetic returns have true volatility $\sigma^{\text{tar}}$ by construction, the distribution of $\sigma^\text{synth}_k$ characterizes the sampling variability we would observe if our index were tracking the target perfectly. This distribution provides the confidence bands shown in Figure \ref{f-vol}.

\section{Closed-Form Approximation for EWMA Estimators}\label{ap-example}
While the Monte Carlo approach above is general, we can derive a closed-form approximation when using EWMA-based volatility estimates. We first consider the simpler case of a simple moving average (SMA) estimator, then extend to EWMA.

For an SMA estimator with window size $m$, the squared volatility estimate from i.i.d.\ normal returns with true standard deviation $\sigma$ is
\[
\left(\sigma^\text{SMA}\right)^2 = \frac{1}{m}\sum_{i=1}^{m} \left(r_i\right)^2.
\]
Since each $r_i \sim \mathcal{N}(0, \sigma^2)$, we have $(r_i/\sigma)^2 \sim \chi^2_1$, and the sum of $m$ independent $\chi^2_1$ variables follows a $\chi^2_m$ distribution. Thus $(\sigma^\text{SMA})^2 \sim (\sigma^2/m) \chi^2_m$, or equivalently, $\sigma^\text{SMA}$ follows a scaled chi distribution with $m$ degrees of freedom and scale $\sigma/\sqrt{m}$.

For an EWMA estimator with decay factor $\beta = \exp(-\log(2)/h)$ corresponding to halflife $h$, the squared volatility estimate is a weighted sum of squared returns. In the infinite-sample limit,
\[\left(\sigma^\text{EWMA}\right)^2=(1-\beta)\sum_{i=1}^\infty \beta^{i-1}r_i^2.\]
This is no longer an exact chi-squared, but we can approximate its distribution by moment matching. Using $\mathbb{E}[r_i^2] = \sigma^2$ and $\text{Var}[r_i^2] = 2\sigma^4$ (since $r_i^2/\sigma^2 \sim \chi^2_1$), we obtain
\begin{align*}
    \mathbb{E}[(\sigma^\text{EWMA})^2]&=\sigma^2, \\
    \text{Var}[(\sigma^\text{EWMA})^2]&=(1-\beta)^2 \frac{2\sigma^4}{1-\beta^2}=\frac{2\sigma^4(1-\beta)}{1+\beta}.
\end{align*}
A scaled chi-squared $(\sigma^2/\nu)\chi^2_\nu$ has mean $\sigma^2$ and variance $2\sigma^4/\nu$. Matching variance gives the effective degrees of freedom $\nu = (1+\beta)/(1-\beta)$. Thus $\sigma^\text{EWMA}$ is approximately distributed as a scaled chi distribution with $(1+\beta)/(1-\beta)$ degrees of freedom and scale $\sigma\sqrt{(1-\beta)/(1+\beta)}$.

\begin{figure}[H]
\centering
\includegraphics[width=\linewidth]{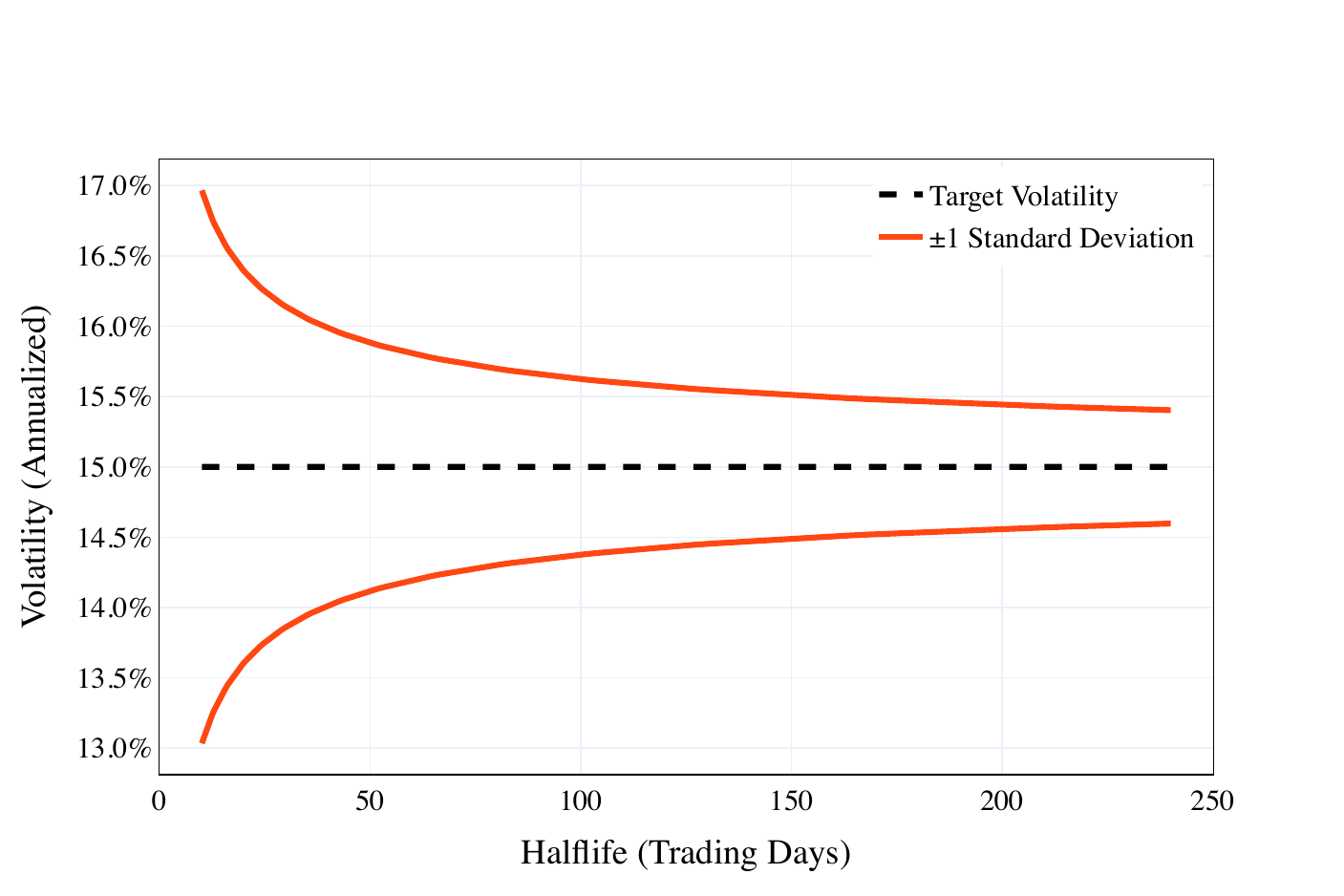}
\caption{Approximate standard deviation of the EWMA volatility estimate as a function of halflife $h$ for $\sigma^{\text{tar}}=0.15/\sqrt{252}$.}
\label{f-vol-std}
\end{figure}

Figure \ref{f-vol-std} shows the approximate standard deviation of $\sigma^\text{EWMA}$ as a function of halflife $h$ for $\sigma^{\text{tar}}=0.15/\sqrt{252}$ (annualized volatility of $15\%$).

\begin{figure}[H]
\centering
\includegraphics[width=\linewidth]{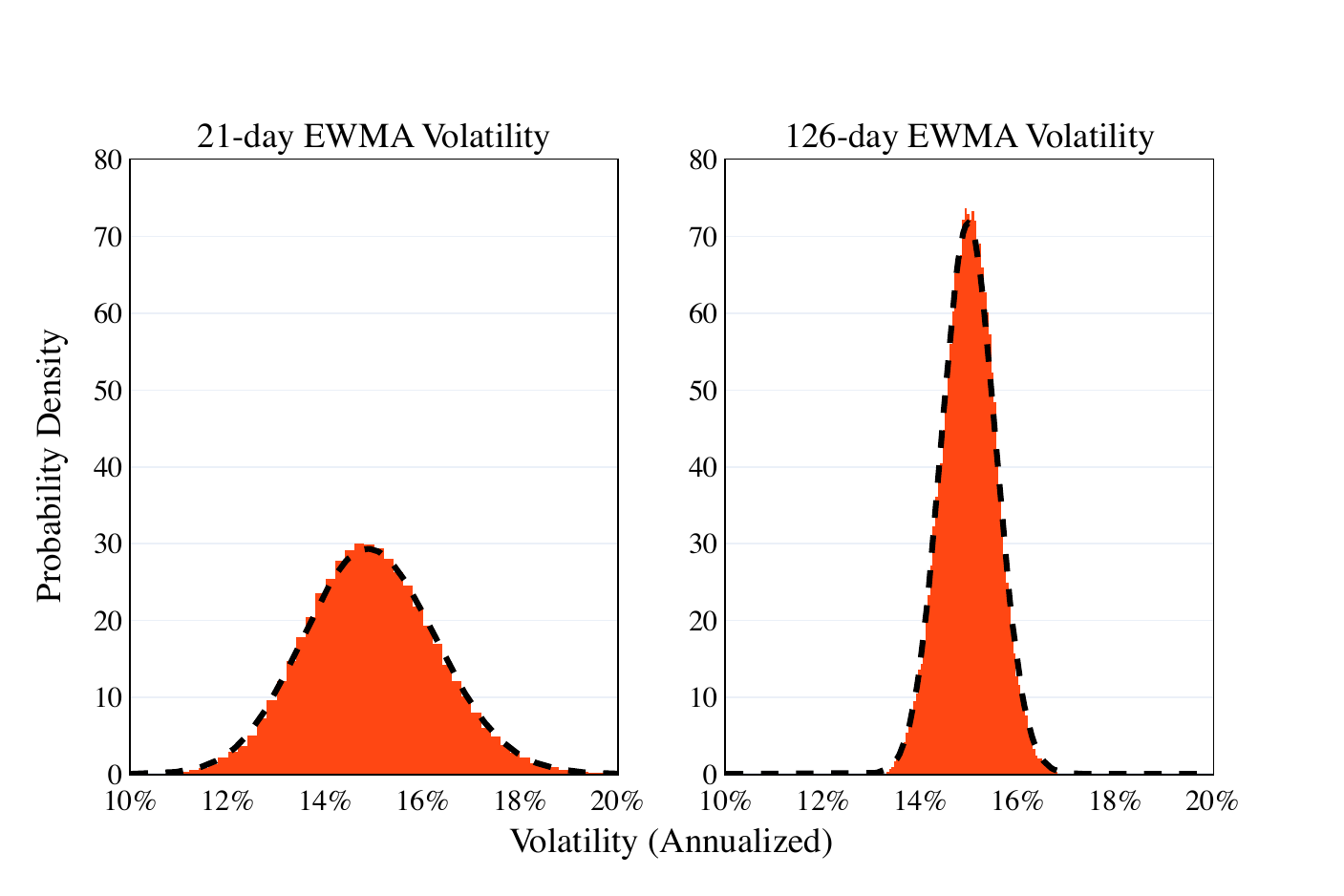}
\caption{Histogram of EWMA volatility estimates from Monte Carlo simulation ($N=10000$, burn-in $n=252$) compared to the chi distribution approximation (dashed lines) for halflives $h=21$ and $h=126$.}
\label{f-vol-mc}
\end{figure}

Figure \ref{f-vol-mc} validates this approximation by comparing the closed-form chi distribution to the empirical distribution from Monte Carlo simulation with $N=10000$ samples.

Figures \ref{f-vol-std} and \ref{f-vol-mc} show that EWMA volatility estimates with longer halflives have significantly less variance.
As such, we recommend using halflives of at least 63 when evaluating the performance of our index construction methods.

\end{document}